\documentclass[aps,prd,twocolumn,superscriptaddress,showpacs]{revtex4}

\usepackage[dvips]{graphicx}
\usepackage{ulem}
\usepackage[usenames]{color}
\usepackage{amsmath}
\usepackage{float}

\begin{document}


\title{Study of $P$-wave excitations of observed charmed strange baryons}


\author{Dan-Dan Ye}
\affiliation{Department of Physics, Shanghai University, Shanghai 200444, China}
\affiliation{College of Mathematics, Physics and Information Engineering, Jiaxing University, Jiaxing 314001, China}

\author{Ze Zhao}
\affiliation{Department of Physics, Shanghai University, Shanghai 200444, China}


\author{Ailin Zhang}
\email{zhangal@staff.shu.edu.cn}
\affiliation{Department of Physics, Shanghai University, Shanghai 200444, China}

\begin{abstract}
Many excited charmed strange baryons such as $\Xi_c(2790)$, $\Xi_c(2815)$, $\Xi_c(2930)$, $\Xi_c(2980)$, $\Xi_c(3055)$, $\Xi_c(3080)$ and $\Xi_c(3123)$ have been observed. In order to understand their internal structure and to determine their spin-parities, the strong decay properties of these baryons as possible $P$-wave excited $\Xi_c$ candidates have been systematically studied in a $^3P_0$ model. The configurations and $J^P$ assignments of $\Xi_c(2790)$, $\Xi_c(2815)$, $\Xi_c(2930)$, $\Xi_c(2980)$, $\Xi_c(3055)$, $\Xi_c(3080)$ and $\Xi_c(3123)$ have been explored based on recent experimental data. In our analyses, $\Xi_c(3055)$, $\Xi_c(3080)$ and $\Xi_c(3123)$ seem impossible to be the $P$-wave excited $\Xi_c$. $\Xi_c(2790)$, $\Xi_c(2815)$, $\Xi_c(2930)$ and $\Xi_c(2980)$ may be the $P$-wave excited $\Xi_c$. In particular, $\Xi_c(2790)$ and $\Xi_c(2815)$ are very possibly the $P$-wave excited $\Xi_{c1}(1/2^-)$ and $\Xi_{c1}(3/2^-)$, respectively.
$\Xi_c(2980)$ may be the $P$-wave excited $\Xi_{c1}^{'}(\frac{1}{2}^-)$. $\Xi_c(2930)$ may be the $P$-wave $\Xi_{c0}^{'}(\frac{1}{2}^-)$, $\tilde{\Xi}_{c0}(\frac{1}{2}^-)$, $\Xi_{c2}^{'}(\frac{3}{2}^-)$, $\Xi_{c2}^{'}(\frac{5}{2}^-)$, $\tilde{\Xi}_{c2}(\frac{3}{2}^-)$ or $\tilde{\Xi}_{c2}(\frac{5}{2}^-)$. Furthermore, some branching fraction ratios related to the internal structure and quark configuration of $P$-wave $\Xi_c$ have also been computed. Measurements of these ratios in the future will be helpful to understand these excited $\Xi_c$.
\end{abstract}

\maketitle

\section{Introduction \label{sec:introduction}}
Quantum chromodynamics (QCD) is the fundamental theory of the strong interaction, which is responsible for the dynamics of hadrons. However, the connection of QCD to hadron is not very clear. People have to understand hadrons through their masses, productions and decays in all kinds of models. The structure and dynamics in baryons are complicated for their three quarks or anti-quarks. Baryons with one $u$ or $d$, one strange and one charmed quark are identified with $\Xi_c$~\cite{pdg,klempt,crede}. A study of $\Xi_c$ is useful for understanding the structure and dynamics in hadron. Except for $\Xi_c(2930)$ and $\Xi_c(3123)$ observed only in one experiment, more and more highly excited charmed strange baryons such as $\Xi_c(2790)$, $\Xi_c(2815)$, $\Xi_c(2980)$, $\Xi_c(3055)$, and $\Xi_c(3080)$ have been observed in different experiments~\cite{pdg}. So far, the spins and parities of these $\Xi_c$ have not yet been measured by experiment, how to determine their spins and parities is an important theoretical topic.

The mass spectra of excited $\Xi_c$ baryons have been computed in many models~\cite{brac,petry,ebert,roberts,guo,huang,zhang1,zhu1,zhang2,zhu2,Zalak Shah:1602.06384 (2016),zhang3}, the strong decays of excited $\Xi_c$ baryons have also been studied in many models~\cite{cheng,zhu3,zhong,zhang4,zhu4,zhang3}. In these models, some possible $J^P$ assignments of these excited $\Xi_c$ have been performed, which is presented in Table~\ref{table1}.

\begin{center}
\begin{table*}[t]
\caption{Some possible $J^P$ assignments of excited $\Xi_c$}
\begin{tabular}{p{0.0cm} p{1.8cm}*{10} {p{1.35cm}}}
\hline\hline
& Resonances & \cite{cheng}&\cite{ebert} & \cite{roberts} & \cite{guo}& \cite{zhang1} &\cite{zhong}&\cite{zhu1} &\cite{zhang2}&\cite{zhu2,zhu4}&\cite{zhang4} \\
   \hline
&$\Xi_c(2790)$ & $\frac{1}{2}^-$ & $\frac{1}{2}^-$ & $\frac{1}{2}^-$ & $\frac{1}{2}^-$  & $\frac{1}{2}^-$ & $\frac{1}{2}^-$ &  $\frac{1}{2}^-$ &$\frac{1}{2}^-$  &   $\frac{1}{2}^-$  &  $\cdots$\\
&$\Xi_c(2815)$ & $\frac{3}{2}^-$ &  $\frac{3}{2}^-$ & $\frac{3}{2}^-$ & $\frac{3}{2}^-$ &$\frac{3}{2}^-$ & $\frac{3}{2}^-$ &  $\frac{3}{2}^-$ & $\frac{3}{2}^-$       & $\frac{3}{2}^-$ & $\cdots$ \\
   &$\Xi_c(2930)$ & $\cdots$    & $(\frac{1}{2},\frac{3}{2},\frac{5}{2})^-$ & $\cdots$   &  $\cdots$   &  $\cdots$ & $\frac{1}{2}^-$ &  $\cdots$    &  $\cdots$       &   $\frac{1}{2}^-$          &  $\cdots$\\
   &$\Xi_c(2980)$ & $\frac{1}{2}^+$     & $\frac{1}{2}^{+*}$ & 4-p   & $\frac{1}{2}^{+*}$   &   $\frac{3}{2}^+$ & $(\frac{1}{2},\frac{3}{2})^-$ &  $(\frac{1}{2},\frac{3}{2})^-$ &  $\frac{1}{2}^{+*}$  &   $\frac{1}{2}^-$     &  $\cdots$\\
   &$\Xi_c(3055)$ & $\cdots$     &   $\frac{3}{2}^+$ & 4-p         & $\frac{5}{2}^+$   &  $\cdots$  & $\frac{3}{2}^+$ &  $\cdots$  &  $\frac{3}{2}^+$  &   $\frac{3}{2}^+$   & $(\frac{5}{2},\frac{7}{2})^+$\\
   &$\Xi_c(3080)$ & $\frac{5}{2}^+$     &   $\frac{5}{2}^+$   & 4-p      & $\frac{5}{2}^+$   &  $\frac{5}{2}^+$ & $\frac{1}{2}^{+*}$ &  $\frac{5}{2}^-$    &  $\frac{5}{2}^+$   &   $\frac{5}{2}^+$    & $\neq$D-wave\\
   &$\Xi_c(3123)$ & $\cdots$   & $\frac{7}{2}^+$ & 2-p  & $\frac{5}{2}^+$ &  $\cdots$ & $(\frac{3}{2},\frac{5}{2})^+$ & $\cdots$ & $(\frac{1}{2},\frac{3}{2})^{-*}$ & $\cdots$  &  $\cdots$\\
   \hline\hline
\end{tabular}
\label{table1}
\end{table*}
\end{center}

In the table, $4$-p and $2$-p indicate the $4$ and $2$ possible assignments of $J^P$, respectively, mentioned in table $11$ in Ref.~\cite{roberts}. $(\frac{1}{2},\frac{3}{2},\frac{5}{2})^-$ indicate three $J^P$ possibilities: $\frac{1}{2}^-$, $\frac{3}{2}^-$, and $\frac{5}{2}^-$. The same notation applies to $(\frac{3}{2},\frac{5}{2})^+$ and $(\frac{5}{2},\frac{7}{2})^+$. $\frac{1}{2}^{+*}$ indicates the radially excited $\frac{1}{2}^{+}(2S)$, while $(\frac{1}{2},\frac{3}{2})^{-*}$ indicates the radially excited $\frac{1}{2}^-(2P)$ and $\frac{3}{2}^-(2P)$, respectively.

From this table, the $J^P$ assignments of $\Xi_c(2790)(\frac{1}{2}^-)$ and $\Xi_c(2815)(\frac{3}{2}^-)$ are the same in all references. However, there are different $J^P$ assignments for other excited $\Xi_c(2930)$, $\Xi_c(2980)$, $\Xi_c(3055)$, $\Xi_c(3080)$, and $\Xi_c(3123)$.

$^3P_0$ model as a phenomenological method has been employed successfully to study the Okubo-Zweig-Iizuka(OZI)-allowed hadronic decays of
hadrons~\cite{micu1969,yaouanc1,yaouanc2,yaouanc3}. In Ref.~\cite{zhu3}, the strong decays of $S,P,D-$wave charmed baryons have been systematically studied in the framework of the $^3P_0$ model. For lack of experimental data at that time, only the possible internal structure and quantum numbers were discussed for $\Xi_c(2980)$ and $\Xi_c(3080)$. In Ref.~\cite{zhang4}, $\Xi_c(3055)$ and $\Xi_c(3080)$ were analyzed in the $^3P_0$ model after the report of Belle collaboration~\cite{belle1}. Systematical study of all these excited $\Xi_c$ in the $^3P_0$ model has not yet been performed.

Very recently, the Belle collaboration presented new measurements of the masses and widths of the $\Xi_c^{'}$, $\Xi_c(2645)$, $\Xi_c(2790)$, $\Xi_c(2815)$, and $\Xi_c(2980)$~\cite{belle2}. In particular, some relative branching fractions were also estimated in the experiment as follows:
$$\frac{B(\Xi_{c}(2815)^{+} \to \Xi_c^{'0} \pi^{+})}{B(\Xi_{c}(2815)^{+} \to \Xi_{c}(2645)^{0}\pi^{+},\Xi_{c}(2645)^{0} \to \Xi_c^{+}\pi^{-})}\approx 11\%, $$
$$\frac{B(\Xi_{c}(2815)^{0} \to \Xi_c^{'+} \pi^{-} )}{B(\Xi_{c}(2815)^{0} \to \Xi_{c}(2645)^{+}\pi^{-},\Xi_{c}(2645)^{+} \to \Xi_c^{0}\pi^{+})}\approx 10\%, $$
$$\frac{B(\Xi_{c}(2980)^{+} \to \Xi_c^{'0} \pi^{+} )}{B(\Xi_{c}(2815)^{+} \to \Xi_{c}(2645)^{0}\pi^{+},\Xi_{c}(2645)^{0} \to \Xi_c^{+}\pi^{-})}\approx 75\%, $$
$$\frac{B(\Xi_{c}(2980)^{0} \to \Xi_c^{'+} \pi^{-} )}{B(\Xi_{c}(2815)^{0} \to \Xi_{c}(2645)^{+}\pi^{-},\Xi_{c}(2645)^{+} \to \Xi_c^{0}\pi^{+})}\approx 50\%. $$
These branching fraction ratios provide more information to understand the nature of $\Xi_c(2815)$ and $\Xi_c(2980)$.

In order to have a comprehensive understanding of these excited $\Xi_c$ baryons, the hadronic decay of these $\Xi_c$ baryons as $P$-wave charmed strange candidates will be systematically studied in the $^3P_0$ model in this paper, while other possible assignments of these excited $\Xi_c$ are reserved in another paper.

The paper is organized as follows. In Sec. II, we give a brief review of the $^3P_0$ model. We present our numerical results and analyzes of these excited $\Xi_c$ in Sec. III. In the last section, we give our conclusions and discussions.

\section{Baryon decay in the $^3P_0$ model \label{Sec: $^3P_0$ model}}

As well known, the $^3P_0$ model was first proposed by Micu~\cite{micu1969}, and further developed by Orsay Group~\cite{yaouanc1,yaouanc2,yaouanc3}. The model has been widely employed to study the OZI-allowed strong decays of hadrons by many authors not cited here.

Although the hadronic decay dynamics or the connection between the $^3P_0$ model and QCD is not clear, some attempts to make a bridge between the phenomenological $^3P_0$ model and the more fundamental ingredients of QCD have been made both in meson sector~\cite{alcock,kokoski,geiger,ackleh,roberts,bonnaz,close} and in baryon sector~\cite{capstick,capstick2000}. In the model, a pair of quark $q\bar q$ is assumed to be created from the vacuum then to regroup with the quarks from the initial hadron $A$ to form two daughter hadrons $B$ and $C$. In particular, the interaction Hamiltonian for the production process was assumed as~\cite{geiger,ackleh,close}
\begin{eqnarray}
H_{q \bar q} = \gamma\, \sum_f 2 m_f \int d^{\, 3} x\; \bar \psi_f \psi_f \ ,
\end{eqnarray}
where $\psi_f$ is a Dirac quark field with flavour $f$. $m_f$ is the constituent quark mass,
and $\gamma$ is a dimensionless $q\bar q$ pair-production strength of the decay interaction. The $\gamma$ is often regarded as a free flavor independent constant and is fitted to the data. However, the strength $\gamma$ was regarded as a scale-dependent function of the reduced mass of the decaying meson in Ref.~\cite{segovia2012}.

For the decay of a baryon, three possible rearrangements, namely any of the three quarks in $A$ can go into $C$, are taken into account, which is shown in Fig. 1. Therefore, there are three possibilities in the decay process of a $\Xi_c$ baryon~\cite{zhu3}
\begin{eqnarray}
A(q_1s_2c_3)+P(q_4\overline{q}_5)\to B(q_1q_4s_2)+C(c_3\overline{q}_5), \\
A(q_1s_2c_3)+P(q_4\overline{q}_5)\to B(q_1q_4c_3)+C(s_2\overline{q}_5), \\
A(q_1s_2c_3)+P(q_4\overline{q}_5)\to B(q_4s_2c_3)+C(q_1\overline{q}_5)
\end{eqnarray}
where $q_i$ denotes the $u$ or $d$ quark, while $s_2$ and $c_3$ denote strange and charm quark, respectively.
\begin{figure}
\begin{center}
\includegraphics[height=5.8cm,angle=0,width=8.5cm]{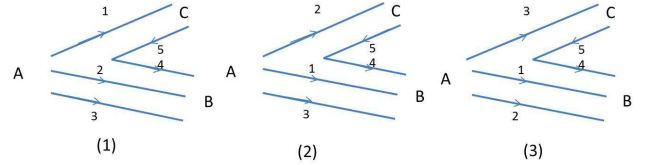}
\caption{Baryon decay process of $A\to B+C$ in the $^3P_0$ model.}
\end{center}
\end{figure}

In the $^3P_0$ model, the hadronic decay width $\Gamma$ of a process $A \to B + C$ is as follows~\cite{yaouanc3},
\begin{eqnarray}
\Gamma  = \pi ^2 \frac{|\vec{p}|}{m_A^2} \frac{1}{2J_A+1}\sum_{M_{J_A}M_{J_B}M_{J_C}} |{\mathcal{M}^{M_{J_A}M_{J_B}M_{J_C}}}|^2.
\end{eqnarray}
In the equation, $\vec{p}$ is the momentum of the daughter baryon in $A$'s center of mass frame,
\begin{eqnarray}
 |\vec{p}|=\frac{{\sqrt {[m_A^2-(m_B-m_C )^2][m_A^2-(m_B+m_C)^2]}}}{{2m_A, }}
\end{eqnarray}
$m_A$ and $J_A$ are the mass and the total angular momentum of the initial baryon $A$, respectively. $m_B$ and $m_C$ are the masses of the final hadrons. $\mathcal{M}^{M_{J_A}M_{J_B}M_{J_C}}$ is the helicity amplitude. For a process of $\Xi_c$ decaying into a charmed baryon and a light meson, the  $\mathcal{M}^{M_{J_A}M_{J_B}M_{J_C}}$ reads~\cite{zhu3,zhang4,zhang5} (there is a factor $2$ difference from that in Ref.~\cite{zhu3} for different initial $A$.)
\begin{flalign}
 &\delta^3(\vec{p}_B+\vec{p}_C-\vec{p}_A)\mathcal{M}^{M_{J_A } M_{J_B } M_{J_C }}\nonumber \\
 &=-\gamma\sqrt {8E_A E_B E_C }  \sum_{M_{\rho_A}}\sum_{M_{L_A}}\sum_{M_{\rho_B}}\sum_{M_{L_B}} \sum_{M_{S_1},M_{S_3},M_{S_4},m}  \nonumber\\
 &\langle {J_{l_A} M_{J_{l_A} } S_3 M_{S_3 } }| {J_A M_{J_A } }\rangle \langle {L_{\rho_A} M_{L_{\rho_A} } L_{\lambda_A} M_{L_{\lambda_A} } }| {L_A M_{L_A } }\rangle \nonumber \\
 &\langle L_A M_{L_A } S_{12} M_{S_{12} }|J_{l_A} M_{J_{l_A}} \rangle \langle S_1 M_{S_1 } S_2 M_{S_2 }|S_{12} M_{S_{12} }\rangle \nonumber \\
 &\langle {J_{l_B} M_{J_{l_B} } S_3 M_{S_3 } }| {J_B M_{J_B } }\rangle \langle {L_{\rho_B} M_{L_{\rho_B} } L_{\lambda_B} M_{L_{\lambda_B} } }| {L_B M_{L_B } }\rangle \nonumber \\
 &\langle L_B M_{L_B } S_{14} M_{S_{14} }|J_{l_B} M_{J_{l_B}} \rangle \langle S_1 M_{S_1 } S_4 M_{S_4 }|S_{14} M_{S_{14} }\rangle \nonumber \\
 &\langle {1m;1 - m}|{00} \rangle \langle S_4 M_{S_4 } S_5 M_{S_5 }|1 -m \rangle \nonumber \\
 &\langle L_C M_{L_C } S_C M_{S_C}|J_C M_{J_C} \rangle \langle S_2 M_{S_2 } S_5 M_{S_5 }|S_C M_{S_C} \rangle \nonumber \\
&\times\langle\varphi _B^{1,4,3} \varphi _C^{2,5}|\varphi _A^{1,2,3}\varphi _0^{4,5} \rangle \times I_{M_{L_B } ,M_{L_C } }^{M_{L_A },m} (\vec{p}),
\end{flalign}
where $\langle\varphi _B^{1,4,3} \varphi _C^{2,5}|\varphi _A^{1,2,3}\varphi _0^{4,5} \rangle$ are the matrix elements of flavor wave functions. $I_{M_{L_B } ,M_{L_C } }^{M_{L_A },m} (\vec{p})$ are the spatial integrals which describe the overlap of the initial baryon and the created two final hadrons

\begin{flalign}
I_{M_{L_B } ,M_{L_C } }^{M_{L_A } ,m} (\vec{p})&= \int d \vec{p}_1 d \vec{p}_2 d \vec{p}_3 d \vec{p}_4 d \vec{p}_5 \nonumber \\
&\times\delta ^3 (\vec{p}_1 + \vec{p}_2 + \vec{p}_3 -\vec{p}_A)\delta ^3 (\vec{p}_4+ \vec{p}_5)\nonumber \\
&\times \delta ^3 (\vec{p}_1 + \vec{p}_4 + \vec{p}_3 -\vec{p}_B )\delta ^3 (\vec{p}_2 + \vec{p}_5 -\vec{p}_C) \nonumber \\
& \times\Psi _{B}^* (\vec{p}_1, \vec{p}_4,\vec{p}_3)\Psi _{C}^* (\vec{p}_2 ,\vec{p}_5) \nonumber \\
& \times \Psi _{A} (\vec{p}_1 ,\vec{p}_2 ,\vec{p}_3)y _{1m}\left(\frac{\vec{p_4}-\vec{p}_5}{2}\right).
\end{flalign}

In principle, the meson wave functions are usually determined in a nonrelativistic quark model with traditional Coulomb $+$ linear and smeared hyperfine interactions. However, as indicated in Ref~\cite{ackleh} ``Experience indicates, however, that it is not useful to employ more accurate Coulomb plus linear wave functions in what must be highly simplified models of complex hadronic interactions". In Refs.~\cite{blundell,close}, it was also pointed out that the $^3P_0$ model and experiment were sufficiently imprecise that computations with more realistic quark model reveal no systematic improvements. In particular, the $^3P_0$ model gives a good description of many of the observed decay amplitudes and partial widths of mesons in terms of the simple harmonic oscillator wave functions. Therefore, the simple harmonic oscillator wave functions have been employed for both the meson and the baryons in previous equation. The baryon wave functions (with definite total $L=L_\rho+L_\lambda$) are often made from a Clebsch-Gordan sum of harmonic oscillator wave functions in the two relative coordinates $\rho$ and $\lambda$~\cite{capstick}. The explicit expressions of the baryon wave functions and the spatial integrals $I_{M_{L_B } ,M_{L_C } }^{M_{L_A },m}$ were presented in Refs.~\cite{zhang4,zhang5}. More details of baryon decays in the $^3P_0$ model could be found in Refs.~\cite{Capstick:1994 (1993),yaouanc3,zhu3,zhang4,zhang5}.

\section{Numerical results \label{Sec: Numerical results}}
\subsection{Notations of baryons and relevant parameters}

In our calculation, notations for the excited baryons are the same as those in Refs.~\cite{cheng,zhu3,zhang4}. In Tables \ref{table1sp}-\ref{table14}, $n_\rho$ and $L_\rho$ denote the nodal and the orbital angular momentum between the two light quarks, $n_\lambda$ and $L_\lambda$ denote the nodal and the orbital angular momentum between the charm quark and the two light quark system, $L$ is the total orbital angular momentum of $L_\rho$ and $L_\lambda$. $S_\rho$ denotes the total spin of the two light quarks, $J_l$ is total angular momentum of $L$ and $S_\rho$. $J$ is the total angular momentum of the baryons. Besides, In Table~\ref{table1sp}, the hat and the check are also used to denote the assignments with $L_\rho=2$ and $L_\rho=1$, respectively. The superscript $L$ is adopted to denote the different total angular momentum in $\check\Xi_{cJ_l}^{\ L}$. For $\Xi_c$ baryons, the relevant quantum numbers of $S,P$-wave of $\Xi_c$ baryons are given in Table.~\ref{table1sp}

\begin{table}[t]
\caption{Quantum numbers of $S,P$-wave of $\Xi_c$ baryons}
\begin{tabular}{p{0.0cm} p{2.5cm}*{6}{p{0.8cm}}}
\hline\hline
             & Assignments                                        & $J$                         & $J_l$ & $L_\rho$ & $L_\lambda$ & $L$  & $S_\rho$ \\
\hline
\label{01}   &$\Xi_{c}^{0,+}(\frac{1}{2}^+)$                      & $\frac{1}{2}$               &  0    &  0       &   0         &  0   &  0       \\
\label{02}   &$\Xi_{c}^{'0,+}(\frac{1}{2}^+)$                     & $\frac{1}{2}$               &  1    &  0       &   0         &  0   &  1       \\
\label{03}   &$\Xi_{c}^{\ast0,+}(\frac{3}{2}^+)$                  & $\frac{3}{2}$               &  1    &  0       &   0         &  0   &  1       \\
\hline\hline
\label{01}   &$\Xi_{c0}^{' }(\frac{1}{2}^-)$                      & $\frac{1}{2}$               &  0    &  0       &   1         &  1   &  1       \\
\label{02}   &$\Xi_{c1}^{' }(\frac{1}{2}^-)$                      & $\frac{1}{2}$               &  1    &  0       &   1         &  1   &  1       \\
\label{03}   &$\Xi_{c1}^{' }(\frac{3}{2}^-)$                      & $\frac{3}{2}$               &  1    &  0       &   1         &  1   &  1       \\
\label{04}   &$\Xi_{c2}^{' }(\frac{3}{2}^-)$                      & $\frac{3}{2}$               &  2    &  0       &   1         &  1   &  1       \\
\label{05}   &$\Xi_{c2}^{' }(\frac{5}{2}^-)$                      & $\frac{5}{2}$               &  2    &  0       &   1         &  1   &  1       \\
\label{06}   &$\Xi_{c1}^{  }(\frac{1}{2}^-)$                      & $\frac{1}{2}$               &  1    &  0       &   1         &  1   &  0       \\
\label{07}   &$\Xi_{c1}^{  }(\frac{3}{2}^-)$                      & $\frac{3}{2}$               &  1    &  0       &   1         &  1   &  0       \\
\label{08}   &$\tilde\Xi_{c1}^{' }(\frac{1}{2}^-)$                & $\frac{1}{2}$               &  1    &  1       &   0         &  1   &  0       \\
\label{09}   &$\tilde\Xi_{c1}^{' }(\frac{3}{2}^-)$                & $\frac{3}{2}$               &  1    &  1       &   0         &  1   &  0       \\
\label{10}   &$\tilde\Xi_{c0}^{' }(\frac{1}{2}^-)$                & $\frac{1}{2}$               &  0    &  1       &   0         &  1   &  1       \\
\label{11}   &$\tilde\Xi_{c1}^{' }(\frac{1}{2}^-)$                & $\frac{1}{2}$               &  1    &  1       &   0         &  1   &  1       \\
\label{12}   &$\tilde\Xi_{c1}^{' }(\frac{3}{2}^-)$                & $\frac{3}{2}$               &  1    &  1       &   0         &  1   &  1       \\
\label{13}   &$\tilde\Xi_{c2}^{' }(\frac{3}{2}^-)$                & $\frac{3}{2}$               &  2    &  1       &   0         &  1   &  1       \\
\label{14}   &$\tilde\Xi_{c2}^{' }(\frac{5}{2}^-)$                & $\frac{5}{2}$               &  2    &  1       &   0         &  1   &  1       \\
\hline\hline
\end{tabular}
\label{table1sp}
\end{table}

Masses of relevant mesons and baryons involved in our calculation are listed in Table~\ref{table3}~\cite{pdg}. The main model parameters involved in our calculation are the pair-creation strength $\gamma$, the $R$ in the harmonic oscillator wave functions of the meson and the $\beta_{\lambda,\rho}$ in the baryon wave functions, respectively. The $R$ is taken to be $2.5$ GeV$^{-1}$ for $\pi/K$ meson and $1.67$ GeV$^{-1}$ for $D$ meson as those in Refs.~\cite{S Godfrey1,S Godfrey2}. Different $\beta$ and $\gamma$ have been employed for baryons in different references~\cite{N Isgur,P Geiger,I D¡¯Souza:0101141,zhu3,Capstick:1994 (1993),R Bijker:1506.07469,chenbing:1609.07967}. In this paper, we let $\beta$ to vary in a range from $0.25$ GeV to $0.40$ GeV. In other words, $\beta_{\lambda,\rho}= 0.25\sim0.4$ GeV.

To fit the experimental data for different $\gamma$ and $\beta_{\lambda,\rho}$, the mass and decay widths of $\Xi_c(2645)^{+}$~\cite{pdg} will be taken as a benchmark in the fitting process. In the end, $\gamma=10.5$ and $\gamma=13.6$ were fixed for $\beta_{\lambda,\rho}=250$ MeV and $\beta_{\lambda,\rho}=400$ MeV, respectively. These two groups of $\{\gamma, \beta_{\lambda,\rho}\}$, $\{10.5, 250~\rm {MeV}\}$ and $\{13.4, 400~\rm {MeV}\}$, will be employed in all our calculations. Accordingly, results corresponding to these two groups of parameters are all explicitly presented in the following tables. The partial and total decay widths of $\Xi_c(2645)^{+}$ as the $S$-wave $\Xi_{c1}^{*}(\frac{3}{2}^+)$ were calculated and presented in Table~\ref{table4}. The branching fraction ratios $\Gamma(\Xi_c^{(0/+)}\pi^{(+/-)})/\Gamma_{total}$ of $\Xi_c(2645)^{0}$ related to experiment were also given in the table. The vanish modes in these tables indicate forbidden channels or channels with very small decay width. The ``$\cdots$" in these tables indicates that there exists no such term.

In the calculations, it would be fundamental to take use of the theoretically predicted parameters such as the hadron masses, $\gamma$, $R$ and $\beta_{\lambda,\rho}$ involved in the model. Unfortunately, these inputs for baryons have not been theoretically determined. In addition to previous choice of $\gamma$, $R$ and $\beta_{\lambda,\rho}$, we employed the experimentally measured masses of both baryons and mesons as our inputs in Table.~\ref{table3}. In fact, the energies predicted by the model and the particular parameters involved do not necessarily correlate with the experimentally observed energies.

\begin{table}
\caption{Masses of mesons and baryons involved in the decays~\cite{pdg}}
\begin{tabular}{p{0.0cm} p{2.0cm}p{2.0cm}|p{2.0cm}p{2.0cm}}
   \hline\hline
   &State              &Mass (MeV)  & State          &Mass (MeV)\\
   \hline
   &$\pi^{\pm}      $ &139.570    &$\Xi_c^0$                &2470.85  \\
   &$\pi^{0}        $ &134.977    &$\Xi_c^+$                &2467.93  \\
   &$K^{\pm}        $ &493.677    &$\Sigma_c^{++}$          &2453.97 \\
   &$K^{0}          $ &497.611    &$\Sigma_c^{+} $          &2452.9  \\
   &$\Lambda_c^{+}$ & 2286.46    &$\Sigma_c^{0} $          &2453.75  \\
   &$D^{0}$ & 1864.84 & $\Sigma_c(2520)^{++}$ & 2518.41  \\
   & $D^{+} $  & 1869.59 & $\Sigma_c(2520)^{+}$ & 2517.5  \\
   & $\Lambda$  & 1115.68 & $\Sigma_c(2520)^{0}$ & 2518.48 \\
   & $\Xi_c(3055)^+  $ & 3055.9     & $\Xi_c(3080)^+$   & 3077.2   \\
   &$\Xi_c(2930)    $ &2931       &$\Xi_c(3123)$            &3122.9  \\

   \hline\hline
\end{tabular}
\label{table3}
\end{table}

\begin{center}
  \begin{table*}
\caption{Decay widths (MeV) of $\Xi_c(2645)^+$ and $\Xi_c(2645)^0$. $\mathcal{B}=\Gamma(\Xi_c^{(0/+)}\pi^{(+/-)})/\Gamma_{total}$. The ranges stand for the results with parameters $\{\gamma, \beta_{\lambda,\rho}\}$ from $\{10.5, 250~\rm {MeV}\}$ to $\{13.4, 400~\rm {MeV}\}$.}
\begin{tabular}{ p{0.0cm} c| ccccc | cccccccccccccccccccc}

   \hline\hline
   & Resonance              & $ n_{\lambda} $ & $L_\lambda$  & $ n_{\rho} $ & $L_\rho$   & $S_\rho$
   & $\Xi_c^0\pi^{(+/0)} $    & $\Xi_c^{+}\pi^{(0/-)} $     & $\Gamma_{total}$     & $\Gamma_{exp}$    & $\mathcal{B}$ \\
   \hline
  &$ \Xi_c(2645)^+ (\frac{3}{2}^+)$
  & 0  &  0  &  0  & 0 &  1   &$1.46\sim1.45$ & $0.72\sim0.71$    & input           &$ 2.14\pm0.19$    &$0.67\sim 0.67$\\
  &$ \Xi_c(2645)^0 (\frac{3}{2}^+)$
  & 0  &  0  &  0  & 0 &  1   &$0.76\sim0.75$ & $1.50\sim1.49$    & $2.26\sim2.24$  &$ 2.35\pm0.22$     &$0.66\sim0.66$\\

   \hline\hline
\end{tabular}
\label{table4}
\end{table*}
\end{center}

\subsection{Decays of $\Xi_c(2790)$, $\Xi_c(2815)$ and $\Xi_c(2980)$ }

The $\Xi_c(2790)$ was found in $\Xi_c^{'}\pi$ mass spectrum, while the $\Xi_c(2815)$ was found in the $\Xi_c(2645)\pi$ mode. Recently, the measurements of their intrinsic widths were reported by the Belle collaboration~\cite{belle2}: the total decay width  $\Gamma_{\Xi_c(2790)^+ }=8.9\pm0.6\pm0.8 $ MeV and $\Gamma_{\Xi_c(2815)^+ }=2.43\pm0.20\pm0.17$ MeV. In addition, the ratio of branching fraction, $B(\Xi_{c}(2815)^{+} \to \Xi_c^{'0} \pi^{+})/B(\Xi_{c}(2815)^{+} \to \Xi_{c}(2645)^{0}\pi^{+},\Xi_{c}(2645)^{0} \to \Xi_c^{+}\pi^{-})\approx 11\%$, was also reported in Ref.~\cite{belle2}.

The $\Xi_c(2790)$ and $\Xi_c(2815)$ were assigned with the $P$-wave excited $J^P=\frac{1}{2}^-$ and $J^P =\frac{3}{2}^-$ $\Xi_c$, respectively, by the masses, decay modes, or relevant decay widths in existed references. In a constituent quark model, there are six possibilities to describe the excited $\Xi_c(2790)$ and $\Xi_c(2815)$ baryons. In the six possibilities, three possibilities are $\lambda$-mode excitations while the other three are $\rho$-mode excitations. These $\lambda$ and $\rho$-mode excitations are presented in Tables~\ref{table5} and \ref{table6}. As $J^P=\frac{1}{2}^-$ and $J^P =\frac{3}{2}^-$ $P$-wave candidates of $\Xi_c$, possible decay modes and corresponding hadronic decay widths of $\Xi_c(2790)$ and $\Xi_c(2815)$ have been computed and shown in the tables. The vanish modes in these tables indicate forbidden channels or channels with very small decay width. Some ratios of branching fraction of $\Xi_c(2815)$ related to experiments are particularly given. In comparison with experiments, it is reasonable to assign $\Xi_c(2790)$ and $\Xi_c(2815)$ with the $J^P=\frac{1}{2}^-$ and $J^P=\frac{3}{2}^-$ $P$-wave $\Xi_c$, respectively.

\begin{center}
  \begin{table*}[htbp]
\caption{Decay widths (MeV) of $\Xi_c(2790)^+$ as a $J^P=\frac{1}{2}^-$ $P$-wave candidate. The ranges stand for the results with parameters $\{\gamma, \beta_{\lambda,\rho}\}$ from $\{10.5, 250~\rm {MeV}\}$ to $\{13.4, 400~\rm {MeV}\}$.}
\begin{tabular}{cc| ccccc | cccccccccccccccccccccccccccccc}

 \hline\hline
  &$ \Xi_{cJ_l} (J^P) $ & $ n_{\lambda} $ & $L_\lambda$  & $ n_{\rho} $ & $L_\rho$   & $S_\rho$
    &$\Xi_c^0\pi^+    $ &$\Xi_c^{'0}\pi^+ $ &$\Xi_c^{*0}\pi^+$ &$\Xi_c^{+}\pi^0 $  &$\Xi_c^{'+}\pi^0 $ &$\Xi_c^{*+}\pi^0 $
    &$\Lambda_c^{+}K^0$  &$\Gamma_{total}$  & $$ \\
   \hline

  &$\Xi_{c0}^{'}(\frac{1}{2}^-)$  & 0 &  1 &  0 & 0  &  1
  &$16.2\sim59.1$  &0             &0       &$7.9\sim29.8$  &0              &0     &$20.3\sim40.2$  &$44.5\sim129.1$  &  \\
  &$\Xi_{c1}^{'}(\frac{1}{2}^-)$  & 0 &  1 &  0 & 0  &  1   &0   &$9.6\sim21.9$ &0.0     &0  &$4.9\sim11.2$  &0.0   &0     &$14.5\sim33.1$   &  \\
  &$\Xi_{c1}(\frac{1}{2}^-)$      & 0 &  1 &  0 & 0  &  0  &0    &$4.8\sim10.9$ &0.0     &0   &$2.4\sim5.6$   &0.0   &0  &$7.2\sim16.6$    &  \\
  &$\tilde{\Xi}_{c1}^{'}(\frac{1}{2}^-)$& 0&0&0 & 1  &  0  &0    &$14.4\sim32.8$&0.0     &0   &$7.3\sim16.8$  &0.0   &0   &$21.7\sim49.7$   &  \\
  &$\tilde{\Xi}_{c0}(\frac{1}{2}^-)$& 0 &0 &  0 & 1  &  1
  &$48.7\sim177.4$ &0             &0       &$23.7\sim89.3$ &0              &0     &$61.0\sim121$   &$133\sim 387$ &  \\
  &$\tilde{\Xi}_{c1}(\frac{1}{2}^-)$& 0 &0 &  0 & 1  &  1
  &0               &$28.9\sim65.7$&0.0     &0              &$14.6\sim33.6$ &0.0   &0               &$43.4\sim 99.3$  & \\
\hline\hline
\end{tabular}
\label{table5}
\end{table*}
\end{center}

\begin{center}
\centering
  \begin{table*}[htbp]
\caption{Decay widths (MeV) of $\Xi_c(2815)^+$ as a $J^P=\frac{3}{2}^-$ $P$-wave candidate. $\mathcal{B}_1 =B(\Xi_{c}(2815)^{+} \to \Xi_c^{'0} \pi^{+})/B(\Xi_{c}(2815)^{+} \to \Xi_{c}(2645)^{0}\pi^{+})$; $\mathcal{B}_1^{'} =B(\Xi_{c}(2815)^{+} \to \Xi_c^{'0} \pi^{+})/B(\Xi_{c}(2815)^{+} \to \Xi_{c}(2645)^{0}\pi^{+},\Xi_{c}(2645)^{0} \to \Xi_c^{+}\pi^{-})$. The ranges stand for the results with parameters $\{\gamma, \beta_{\lambda,\rho}\}$ from $\{10.5, 250~\rm {MeV}\}$ to $\{13.4, 400~\rm {MeV}\}$.}
\begin{tabular}{cc| ccccc | cccccccccccccccccccccccccccccc}

   \hline\hline
   &$ \Xi_{cJ_l}(J^P)  $ & $ n_{\lambda} $ & $L_\lambda$  & $ n_{\rho} $ & $L_\rho$   & $S_\rho$
    &$\Xi_c\pi    $ &$\Xi_c^{'}\pi$ &$\Xi_c^{*}\pi$ &$\Lambda_c^{+}K^0$  &$\Gamma_{total}$  & $\mathcal{B}_1$ & $\mathcal{B}_1^{'} $\\
   \hline

  &$\Xi_{c1}^{'}(\frac{3}{2}^-)$  & 0 &  1 &  0 & 0  &  1  &0            &$0.1\sim0.05$  &$9.2\sim18.7$  &0               &$9.3\sim18.7$  &$0.01\sim0.3 $ &$1.8\sim0.4 \%$\\
  &$\Xi_{c2}^{'}(\frac{3}{2}^-)$  & 0 &  1 &  0 & 0  &  1  &$2.9\sim1.7$ &$0.2\sim0.1$   &0.0            &$0.2\sim0.1$   &$3.3\sim1.9$   &$33\sim35$     &$4925\sim5344 \%$\\
  &$\Xi_{c1}(\frac{3}{2}^-)$      & 0 &  1 &  0 & 0  &  0  &0            &$0.2\sim0.1$   &$4.6\sim9.3$   &0               &$4.8\sim9.4$   &$0.05\sim0.01$ &$7.2\sim1.7 \%$\\
  &$\tilde{\Xi}_{c1}^{'}(\frac{3}{2}^-)$& 0&0&0 & 1  &  0  &0            &$0.7\sim0.3$   &$13.8\sim28.0$ &0               &$14.5\sim28.3$ &$0.05\sim0.0$  &$7.2\sim1.7 \%$\\
  &$\tilde{\Xi}_{c1}(\frac{3}{2}^-)$& 0 &0 &  0 & 1  &  1  &0            &$0.3\sim0.2$   &$27.6\sim56.0$ &0               &$28.0\sim56.2$ &$0.01\sim0.3$  &$1.8\sim0.4 \%$\\
  &$\tilde{\Xi}_{c2}(\frac{3}{2}^-)$& 0 &0 &  0 & 1  &  1  &$8.1\sim5.0$ &$0.6\sim0.3$   &$0.02\sim0.0$  &$0.6\sim0.2$    &$10.1\sim5.6$  &$33\sim35$     &$4925\sim5344$ \%\\

\hline\hline
\end{tabular}
\label{table6}
\end{table*}
\end{center}
Based on the spectrum analysis below and hadronic decay features from the two tables above, $\Xi_c(2790)$ is very possibly the $\Xi_{c1}(\frac{1}{2}^-)$, and $\Xi_c(2815)$ is very possibly the $\Xi_{c1}(\frac{3}{2}^-)$. In these assignments, the predicted total width $\Gamma_{total}=7.2\sim16.6$ MeV of $\Xi_c(2790)$ is well consistent with the experimental measurement, the predicted total width $\Gamma_{total}=4.8\sim9.4$ MeV of $\Xi_c(2815)$ is about two times larger than the new measurements. Recently, a prediction of the hadronic decay width of $\Xi_c(2815)$ is $\Gamma=7.1$ MeV in Ref.~\cite{cheng} and $\Gamma=4.24$ MeV in Ref.~\cite{chenbing:1609.07967}.
The predicted ratio $\mathcal{B}^{'}_1=B(\Xi_{c}(2815)^{+} \to \Xi_c^{'0} \pi^{+})/B(\Xi_{c}(2815)^{+} \to \Xi_{c}(2645)^{0}\pi^{+},\Xi_{c}(2645)^{0} \to \Xi_c^{+}\pi^{-})=7.2\sim1.7\%$ is compatible with the measured $11\%$.

From Tables~\ref{table5} and \ref{table6}, it is difficult to distinguish the $\Xi_{c1}(\frac{3}{2}^-)$ and the $\tilde{\Xi}_{c1}^{'}(\frac{3}{2}^-)$ of $\Xi_c(2815)^+$ from each other through their hadronic decay features for theoretical and experimental uncertainties. It is also difficult to distinguish the $\Xi_{c1}(\frac{1}{2}^-)$ and the $\tilde{\Xi}_{c1}^{'}(\frac{1}{2}^-)$ of $\Xi_c(2790)^+$ from each other through their hadronic decay features. On the other hand, the difficulty implies complicated internal structures of baryons.

However, if a simple potential model based on the harmonic oscillator as shown in Ref.~\cite{klempt} is employed, the energy levels for baryons with one heavy quark (mass $M$) and two light quarks (mass $m$) are described by
\begin{eqnarray}
 E=\sqrt{K/m}(3+2L_{\rho}+4n_{\rho})+\sqrt{K/\mu}(3+2L_{\lambda}+4n_{\lambda})
\end{eqnarray}
where $K$ is a constant describing the potential and $\mu^{-1}=(2M^{-1}+m^{-1})/3 $. Obviously, the $\lambda$ excitations are lower than their $ \rho$ analogs. Therefore the lowest-lying excitations are those with $(n_{\lambda}, L_{\lambda},n_{\rho},L_{\rho})=(0,1,0,0)$. Therefore, $\Xi_c(2790)^+$ and $\Xi_c(2815)^+$ can be well classified as  $\Xi_{c1}(\frac{1}{2}^-)$ and  $\Xi_{c1}(\frac{3}{2}^-)$, respectively.
In the following, the branching fraction ratios $\Gamma(\Xi_c(2645)^{(0)}\pi^{(+)})/\Gamma_{total}$ of $\Xi_c(2815)$ as $\Xi_{c1}(\frac{3}{2}^-)$ are used to calculate the ratios $\mathcal{B}_2^{'}=B(\Xi_{c}(2980)^{+} \to \Xi_c^{'0} \pi^{+})/B(\Xi_{c}(2815)^{+} \to \Xi_{c}(2645)^{0}\pi^{+},\Xi_{c}(2645)^{0} \to \Xi_c^{+}\pi^{-})$ related to experiments.

The $\Xi_c(2980)^+$ state with a width of $43.5\pm7.5\pm7.0$ MeV was first observed by the Belle collaboration in the $\Lambda_c^+ K^- \pi^+ $ channel~\cite{Chistov:0606051} and confirmed by BaBar in the intermediate resonant mode $\Sigma_c(2455)^+ K^-$~\cite{Aubert:0710.5763}. It was also subsequently observed by Belle with a narrower width of $18\pm6\pm3$ MeV in the $\Xi_c(2645)^0\pi^+$ decay channel~\cite{Belle:0802.3968}. In Ref.~\cite{belle2}, a new measurement of $28.1\pm2.4^{+1.0}_{-5.0}$ MeV was presented and the ratio of branching fraction, $ B(\Xi_{c}(2980)^{+} \to \Xi_c^{'0} \pi^{+})/B(\Xi_{c}(2815)^{+} \to \Xi_{c}(2645)^{0}\pi^{+},\Xi_{c}(2645)^{0} \to \Xi_c^{+}\pi^{-})\approx 75\%$, was given. The decay of $\Xi_c(2980)^+$ into $\Lambda_c K $ or $ \Xi_c \pi$ mode has not been observed in experiment.

If $\Xi_c(2980)^+$ is the $P$-wave excitations in $\lambda$- or $ \rho$- mode with negative parity, the $\{(n_\lambda ,L_\lambda ),(n_\rho ,L_ \rho )\}$ =$\{(0,1),(0,0)\}$ or $ \{(0,0),(0,1)\}$. In these assignments, possible decay modes and corresponding hadron decay widths of $\Xi_c(2980)^+$ have been presented in Table~\ref{table7}.

\begin{center}
  \begin{table*}[htbp]
\caption{Decay widths (MeV) of $\Xi_c(2980)^+$ as the $P$-wave excitations. $\mathcal{B}_2^{'}=B(\Xi_{c}(2980)^{+} \to \Xi_c^{'0} \pi^{+})/B(\Xi_{c}(2815)^{+} \to \Xi_{c}(2645)^{0}\pi^{+},\Xi_{c}(2645)^{0} \to \Xi_c^{+}\pi^{-})$; $\mathcal{B}_2=B(\Xi_{c}(2980)^{+} \to \Xi_c^{'0} \pi^{+})/B(\Xi_{c}(2980)^{+} \to \Xi_{c}(2645)^{0}\pi^{+})$; $\mathcal{B}_3=B(\Xi_{c}(2980)^{+} \to \Sigma_c^{++} K^-)/B(\Xi_{c}(2980)^{+} \to \Xi_{c}(2645)^{0}\pi^{+})$. The ranges stand for the results with parameters $\{\gamma, \beta_{\lambda,\rho}\}$ from $\{10.5, 250~\rm {MeV}\}$ to $\{13.4, 400~\rm {MeV}\}$.}
\begin{tabular}{c| ccccc | cccccccccccccccccccccccccccccc}

   \hline\hline
   $ \Xi_{cJ_l} (J^P) $ & $ n_{\lambda} $ & $L_\lambda$  & $ n_{\rho} $ & $L_\rho$   & $S_\rho$
   &$\Xi_c\pi $     &$\Xi_c^{'}\pi $      &$\Xi_c^{*}\pi$      &$\Sigma_c K $  &$\Lambda_c K $
   &$\Gamma_{total}$  & $\mathcal{B}_2^{'}$  & $\mathcal{B}_2 $  & $\mathcal{B}_3$  \\
   \hline

   $\Xi_{c0}^{'}(\frac{1}{2}^-)$   & 0 &  1 &  0 & 0  &  1  &$2.8\sim84.2$  &0              &0             &0              &$5.5\sim83$    &$8.3\sim168$     & $0$ & $\cdots$ & $\cdots$\\
   $\Xi_{c1}^{'}(\frac{1}{2}^-)$   & 0 &  1 &  0 & 0  &  1  &0              &$11\sim65$     &$1.6\sim0.9$  &$56\sim119$     &0              &$69\sim185$    &  $45\sim81\%$ &$7.0\sim77$       &$36\sim144$\\
   $\Xi_{c1}^{'}(\frac{3}{2}^-)$   & 0 &  1 &  0 & 0  &  1  &0              &$2.2\sim1.3$   &$17\sim60$    &$0.06\sim0.03$ &0              &$19\sim61$    &  $18\sim3\%$  &$0.1\sim0.02$     &$0.0$\\
   $\Xi_{c2}^{'}(\frac{3}{2}^-)$   & 0 &  1 &  0 & 0  &  1  &$15\sim12$     &$3.9\sim2.4$   &$1.4\sim0.8$  &$0.1\sim0.05$ &$10.6\sim7.7$   &$31\sim23$    &  $24\sim20\%$ &$2.7\sim3.1$      &$0.08\sim0.07$\\
   $\Xi_{c2}^{'}(\frac{5}{2}^-)$   & 0 &  1 &  0 & 0  &  1  &$15\sim12$     &$1.8\sim1.1$   &$2.2\sim1.2$  &$0.05\sim0.02$ &$10.6\sim7.7$  &$29\sim22$    &  $12\sim9\%$  &$0.8\sim0.9$      &$0.02\sim0.02$\\
   $\Xi_{c1}(\frac{1}{2}^-)$       & 0 &  1 &  0 & 0  &  0  &0              &$5.5\sim33$    &$3.2\sim1.7$  &$28\sim60$     &0              &$37\sim94$    &  $39\sim79\%$ &$1.8\sim19$       &$9.0\sim36$\\
   $\Xi_{c1}(\frac{3}{2}^-)$       & 0 &  1 &  0 & 0  &  0  &0              &$4.4\sim2.7$   &$9.7\sim31$   &$0.1\sim0.05$ &0               &$14\sim33$    &  $49\sim12\%$ &$0.4\sim0.09$     &$0.01\sim0.0$\\

$\tilde{\Xi}_{c1}^{'}(\frac{1}{2}^-)$&0&0   &0   & 1  &  0  &0              &$17\sim98$     &$9.6\sim5.2$  &$85\sim179$     &0              &$111\sim282$   &  $39\sim79\%$    &$1.8\sim19$       &$9.0\sim36$\\
$\tilde{\Xi}_{c1}^{'}(\frac{3}{2}^-)$&0&0   &0   & 1  &  0  &0              &$13\sim8.1$    &$29\sim92$    &$0.3\sim0.15$  &0              &$43\sim100$   &  $49\sim12\%$    &$0.4\sim0.09$     &$0.01\sim0.0$\\
$\tilde{\Xi}_{c0}(\frac{1}{2}^-)$  & 0 &  0 &  0 & 1  &  1  &$8.3\sim252$   &0              &0             &0              &$16.5\sim250$  &$25\sim503$   &  $0$
& $\cdots$ & $\cdots$\\
$\tilde{\Xi}_{c1}(\frac{1}{2}^-)$  & 0 &  0 &  0 & 1  &  1  &0              &$33\sim195$    &$4.8\sim2.6$  &$169\sim358$   &0              &$207\sim556$  &  $44\sim81\%$    &$7.0\sim77$       &$36\sim144$\\
$\tilde{\Xi}_{c1}(\frac{3}{2}^-)$  & 0 &  0 &  0 & 1  &  1  &0              &$6.6\sim4.0$   &$51\sim180$   &$0.2\sim0.08$  &0              &$58\sim184$   &  $18\sim3\%$     &$0.1\sim0.02$     &$0.0$\\
$\tilde{\Xi}_{c2}(\frac{3}{2}^-)$  & 0 &  0 &  0 & 1  &  1  &$45\sim36$     &$12\sim7.3$    &$4.3\sim2.3$  &$0.3\sim0.14$  &$32\sim23$     &$93\sim68$   &  $24\sim20\%$    &$2.7\sim3.1$      &$0.08\sim0.07$\\
$\tilde{\Xi}_{c2}(\frac{5}{2}^-)$  & 0 &  0 &  0 & 1  &  1  &$45\sim36$     &$5.3\sim3.2$   &$6.7\sim3.6$  &$0.1\sim0.06$  &$32\sim23$     &$89\sim66$   &  $12\sim9\%$     &$0.8\sim0.9$      &$0.02\sim0.02$\\

\hline\hline
\end{tabular}
\label{table7}
\end{table*}
\end{center}

From this table, the assignment of $\Xi_c(2980)^+$ with the $\Xi_{c0}^{'}(\frac{1}{2}^-)$, $\Xi_{c2}^{'}(\frac{3}{2}^-)$, $\Xi_{c2}^{'}(\frac{5}{2}^-)$, $\tilde{\Xi}_{c0}(\frac{1}{2}^-)$, $\tilde{\Xi}_{c2}(\frac{3}{2}^-)$ or $\tilde{\Xi}_{c2}(\frac{5}{2}^-)$ seems excluded for a large decay into $\Lambda_c K$ and $\Xi_c \pi$ modes, which disagrees with the experiment. The assignment of $\Xi_c(2980)^+$ with the $\Xi_{c1}^{'}(\frac{3}{2}^-)$, $\Xi_{c1}(\frac{3}{2}^-)$, $\tilde{\Xi}_{c1}^{'}(\frac{3}{2}^-)$ or $\tilde{\Xi}_{c1}(\frac{3}{2}^-)$ seems impossible either for the tiny partial decay width in the $\Sigma_c(2455)^+ K^-$ mode. The assignment of $\Xi_c(2980)$ with the $\tilde{\Xi}_{c1}(\frac{1}{2}^-)$ or $\tilde{\Xi}_{c1}^{'}(\frac{1}{2}^-)$ can be excluded either for a large decay width. In comparison with experiments, the $\Xi_c(2980)^+$ as a $P$-wave excitation with negative parity can be identified with the $\Xi_{c1}^{'}(\frac{1}{2}^-)$.

For a better understanding of the internal structures of $P$-wave $\Xi_c(2980)^+$, some ratios such as $\mathcal{B}_2=B(\Xi_{c}(2980)^{+} \to \Xi_c^{'0} \pi^{+})/B(\Xi_{c}(2980)^{+} \to \Xi_{c}(2645)^{0}\pi^{+})$ and $\mathcal{B}_3=B(\Xi_{c}(2980)^{+} \to \Sigma_c^{++} K^-)/B(\Xi_{c}(2980)^{+} \to \Xi_{c}(2645)^{0}\pi^{+})$ were also computed and presented in the table. When $\Xi_c(2980)^+$ is assigned with the $P$-wave $\Xi_{c1}^{'}(\frac{1}{2}^-)$, the predicted $\mathcal{B}_2=7.0\sim77$, and $\mathcal{B}_3=36\sim144$. The measurement of these ratios in the future will be helpful for the understanding of this $\Xi_c$.

\subsection{Decays of $\Xi_c(3055)$ and $\Xi_c(3080)$}

$\Xi_c(3055)$ was first observed by BaBar~\cite{Aubert:0710.5763} and then confirmed by Belle~\cite{belle3}. $\Xi_c(3080)$ was first observed by Belle~\cite{Chistov:0606051} and then confirmed by BaBar~\cite{Aubert:0710.5763}. Recently, $\Xi_c(3055)$ and $\Xi_c(3080)$ were reported by Belle~\cite{belle1} with
$$m_{\Xi_c(3055)^0} = 3059.0 \pm 0.5(stat) \pm 0.6(sys) ~\rm{MeV},$$ $$m_{\Xi_c(3055)^+} = 3055.8 \pm 0.4(stat) \pm 0.2(sys) ~\rm{MeV},$$ $$m_{\Xi_c(3080)^+} = 3079.6 \pm 0.4(stat) \pm 0.1(sys)~\rm{MeV},$$
and total decay widths from observed $\Lambda D$ mode
$$\Gamma_{\Xi_c(3055)^0} = 6.4 \pm 2.1(stat) \pm 1.1(sys)~\rm{MeV},$$ $$\Gamma_{\Xi_c(3055)^+} = 7.0 \pm 1.2(stat) \pm 1.5(sys)~\rm{MeV},$$ $$\Gamma_{\Xi_c(3080)^+} < 6.3~\rm{MeV}.$$

In particular, some ratios of branching fractions were measured, $$\frac{\Gamma_{\Xi_{c}(3055)^{+} \to \Lambda D^{+}}}{\Gamma_{\Xi_{c}(3055)^{+} \to \Sigma_{c}^{++}K^{-}}}=5.09\pm1.01(stat)\pm0.76(sys),$$
$$\frac{\Gamma_{\Xi_{c}(3080)^{+} \to \Lambda D^{+}}}{\Gamma_{\Xi_{c}(3080)^{+} \to \Sigma_{c}^{++}K^{-}}}=1.29\pm0.30(stat)\pm0.15(sys),$$
$$\frac{\Gamma_{\Xi_{c}(3080)^{+} \to \Sigma_{c}(2520)^{++}K^{-} }}{\Gamma_{\Xi_{c}(3080)^{+} \to \Sigma_{c}^{++}K^{-}}}=1.07\pm0.27(stat)\pm0.04(sys).$$

As a possible candidate of $P$-wave $\Xi_c$, the internal quantum numbers and decay widths of $\Xi_c(3055)$ were presented in Table~\ref{table10}. As a possible candidate of $P$-wave $\Xi_c$, the internal quantum numbers and decay widths of $\Xi_c(3080)$ were presented in Table~\ref{table12}.

\begin{center}
  \begin{table*}[htbp]
\caption{Decay widths (MeV) of $ \Xi_c(3055)^+$ as the $P$-wave $\Xi_c$.
$\mathcal{B}_4=B(\Xi_{c}(3055)^{+} \to \Sigma(2520)^{++}K^{-})/B(\Xi_{c}(3055)^{+} \to \Sigma_c^{++}K^{-})$;
$\mathcal{B}_5=B(\Xi_{c}(3055)^{+} \to \Lambda D^{+})/B(\Xi_{c}(3055)^{+} \to \Sigma_c^{++}K^{-})$;
$\Gamma_{exp}=7.8\pm1.9$ MeV   ;
$\mathcal{B}_5^{exp}=5.09\pm1.01\pm0.76$. The ranges stand for the results with parameters $\{\gamma, \beta_{\lambda,\rho}\}$ from $\{10.5, 250~\rm {MeV}\}$ to $\{13.4, 400~\rm {MeV}\}$.}
\begin{tabular}{c| ccccc | cccccccccccccccccccccccccccccc}

   \hline\hline
   $ \Xi_{cJ_l} (J^P) $ &$ n_{\lambda} $  &$L_\lambda$          &$ n_{\rho} $        &$L_\rho$          &$S_\rho$
                        &$\Xi_c\pi $      &$\Xi_c^{'}\pi $      &$\Xi_c^{*}\pi$      &$\Sigma_c K $     &$\Sigma_c^{*} K $ &$\Lambda_c K $   &$\Lambda D $         &$\Gamma_{total}$    &$\mathcal{B}_4$   &$\mathcal{B}_5 $
                        \\

   $\Xi_{c0}^{'}(\frac{1}{2}^-)$   & 0 &  1 &  0 & 0  &  1  &$0.5\sim57$    &0             &0            &0
   &0               &$0.3\sim48$   &$14\sim103$       &$15\sim208$          &$\cdots$ & $\cdots$\\

   $\Xi_{c1}^{'}(\frac{1}{2}^-)$   & 0 &  1 &  0 & 0  &  1  &0              &$3.0\sim59$   &$5.8\sim3.7$ &$44\sim201$                  &$0.7\sim0.4$    &0             &$28\sim207$       &$81\sim471$          &$0$           &$1.0\sim1.5$\\

   $\Xi_{c1}^{'}(\frac{3}{2}^-)$   & 0 &  1 &  0 & 0  &  1  &0              &$5.5\sim4.1$  &$12\sim67$   &$4.5\sim2.6$       &$49\sim151$     &0             &$2.8\sim2.0$      &$73\sim227$          &$11\sim57$    &$0.9\sim1.2$\\

   $\Xi_{c2}^{'}(\frac{3}{2}^-)$   & 0 &  1 &  0 & 0  &  1  &$26\sim26$     &$9.9\sim7.5$  &$5.2\sim3.4$ &$8.1\sim4.7$ &$0.6\sim0.3$    &$21\sim21$    &$0.6\sim0.4$      &$71\sim63$           &$0.1\sim0.1$  &$0.1\sim0.1$\\

   $\Xi_{c2}^{'}(\frac{5}{2}^-)$   & 0 &  1 &  0 & 0  &  1  &$26\sim26$     &$4.4\sim3.3$  &$8.1\sim5.2$ &$3.6\sim2.1$ &$1.0\sim0.5$    &$21\sim21$    &$8.9\sim6.6$      &$73\sim65$           &$0.3\sim0.3$  &$3.6\sim4.6$\\

   $\Xi_{c1}(\frac{1}{2}^-)$       & 0 &  1 &  0 & 0  &  0  &0              &$1.5\sim30$   &$12\sim7.5$  &$22\sim100$        &$1.4\sim0.7$    &0             &$14\sim103$       &$50\sim242$          &$0.1\sim0.0$  &$1.0\sim1.5$\\

   $\Xi_{c1}(\frac{3}{2}^-)$       & 0 &  1 &  0 & 0  &  0  &0              &$11\sim8.3$   &$10\sim36$   &$9.0\sim5.2$       &$25\sim76$      &0             &$5.6\sim4.1$      &$61\sim130$          &$2.7\sim14$   &$0.9\sim1.2$\\

$\tilde{\Xi}_{c1}^{'}(\frac{1}{2}^-)$&0&0   &0   & 1  &  0  &0              &$4.5\sim89$   &$35\sim22$   &$65\sim301$        &$4.1\sim2.2$       &0             &0                 &$109\sim415$         &$0.1\sim0.0$  &0\\

$\tilde{\Xi}_{c1}^{'}(\frac{3}{2}^-)$&0&0   &0   & 1  &  0  &0              &$33\sim25$    &$31\sim109$  &$27\sim16$         &$74\sim227$        &0             &0                 &$165\sim377$         &$2.7\sim14$   &0\\

$\tilde{\Xi}_{c0}(\frac{1}{2}^-)$  & 0 &  0 &  0 & 1  &  1  &$1.4\sim170$   &0             &0            &0                  &0                  &$0.8\sim145$  &0                 &$2.2\sim315$         &$\cdots$ & $\cdots$ \\

$\tilde{\Xi}_{c1}(\frac{1}{2}^-)$  & 0 &  0 &  0 & 1  &  1  &0              &$8.9\sim178$  &$17\sim11$   &$131\sim603$       &$2.1\sim1.1$       &0             &0                 &$159\sim793$         &$0$           &0\\

$\tilde{\Xi}_{c1}(\frac{3}{2}^-)$  & 0 &  0 &  0 & 1  &  1  &0              &$16\sim12$    &$35\sim201$  &$14\sim7.8$        &$146\sim453$       &0             &0                 &$211\sim675$         &$11\sim57$    &$0$\\

$\tilde{\Xi}_{c2}(\frac{3}{2}^-)$  & 0 &  0 &  0 & 1  &  1  &$78\sim78$     &$30\sim22$    &$16\sim10$   &$24\sim14$  &$1.9\sim1.0$       &$63\sim63$    &0                 &$213\sim188$         &$0.1\sim0.1$  &0\\

$\tilde{\Xi}_{c2}(\frac{5}{2}^-)$  & 0 &  0 &  0 & 1  &  1  &$78\sim78$     &$13\sim9.9$   &$24\sim16$   &$11\sim6.2$  &$2.9\sim1.6$       &$63\sim63$    &0                 &$192\sim174$         &$0.3\sim0.3$  &0\\

\hline\hline
\end{tabular}
\label{table10}
\end{table*}
\end{center}


\begin{center}
  \begin{table*}[htbp]
\caption{Decay widths (MeV) of $\Xi_c(3080)^+$ as the $P$-wave $\Xi_c$.
$\mathcal{B}_4=B(\Xi_{c}(3080)^{+} \to \Sigma(2520)^{++}K^{-})/B(\Xi_{c}(3080)^{+} \to \Sigma_c^{++}K^{-})$;
$\mathcal{B}_5=B(\Xi_{c}(3080)^{+} \to \Lambda D^{+})/B(\Xi_{c}(3080)^{+} \to \Sigma_c^{++}K^{-})$;
$\Gamma_{exp}=3.6\pm1.1$ MeV;
$\mathcal{B}_4^{exp}=1.07\pm0.27\pm0.04$;
$\mathcal{B}_5^{exp}=1.29\pm0.3\pm0.15$. The ranges stand for the results with parameters $\{\gamma, \beta_{\lambda,\rho}\}$ from $\{10.5, 250~\rm {MeV}\}$ to $\{13.4, 400~\rm {MeV}\}$.}
\begin{tabular}{c| ccccc | cccccccccccccccccccccccccccccc}

   \hline\hline
   $ \Xi_{cJ_l} (J^P) $ &$ n_{\lambda} $  &$L_\lambda$          &$ n_{\rho} $        &$L_\rho$          &$S_\rho$
                        &$\Xi_c\pi $      &$\Xi_c^{'}\pi $      &$\Xi_c^{*}\pi$      &$\Sigma_c K $     &$\Sigma_c^{*} K $ &$\Lambda_c K $   &$\Lambda D $         &$\Gamma_{total}$    &$\mathcal{B}_4$   &$\mathcal{B}_5 $
                        \\

   $\Xi_{c0}^{'}(\frac{1}{2}^-)$   & 0 &  1 &  0 & 0  &  1  &$1.5\sim49$    &0             &0            &0
   &0               &$1.3\sim40$   &$9.0\sim98$       &$12\sim188$          &$\cdots$ &$\cdots$\\

   $\Xi_{c1}^{'}(\frac{1}{2}^-)$   & 0 &  1 &  0 & 0  &  1  &0              &$1.7\sim56$   &$7.3\sim4.9$ &$33\sim196$                  &$1.8\sim1.0$    &0             &$18\sim197$       &$61\sim455$          &$0.1\sim0.0$  &$0.8\sim1.5$\\

   $\Xi_{c1}^{'}(\frac{3}{2}^-)$   & 0 &  1 &  0 & 0  &  1  &0              &$6.5\sim5.2$  &$10\sim67$   &$6.7\sim4.1$       &$47\sim171$     &0             &$4.6\sim3.8$      &$75\sim251$          &$6.9\sim41$   &$1.0\sim1.3$\\

   $\Xi_{c2}^{'}(\frac{3}{2}^-)$   & 0 &  1 &  0 & 0  &  1  &$29\sim30$     &$12\sim9.3$   &$6.6\sim4.4$ &$12\sim7.5$ &$1.6\sim0.9$    &$23\sim25$    &$0.9\sim0.8$      &$85\sim78$           &$0.1\sim0.1$  &$0.1\sim0.1$\\

   $\Xi_{c2}^{'}(\frac{5}{2}^-)$   & 0 &  1 &  0 & 0  &  1  &$29\sim30$     &$5.2\sim4.1$  &$10\sim6.9$  &$5.4\sim3.3$ &$2.5\sim1.4$    &$23\sim25$    &$15\sim12$        &$90\sim83$           &$0.5\sim0.4$  &$4.1\sim5.4$\\

   $\Xi_{c1}(\frac{1}{2}^-)$       & 0 &  1 &  0 & 0  &  0  &0              &$0.8\sim28$   &$15\sim9.9$  &$16\sim98$        &$3.6\sim2.1$    &0             &$9.0\sim98$       &$44\sim236$          &$0.2\sim0.0$  &$0.8\sim1.5$\\

   $\Xi_{c1}(\frac{3}{2}^-)$       & 0 &  1 &  0 & 0  &  0  &0              &$13\sim10$    &$11\sim37$   &$13\sim8.3$       &$25\sim86$      &0             &$9.2\sim7.5$      &$71\sim149$          &$1.8\sim10$   &$1.0\sim1.3$\\

$\tilde{\Xi}_{c1}^{'}(\frac{1}{2}^-)$&0&0   &0   & 1  &  0  &0              &$2.5\sim84$   &$44\sim30$   &$49\sim294$        &$11\sim6.2$        &0             &0                 &$106\sim414$         &$0.2\sim0.0$  &0\\

$\tilde{\Xi}_{c1}^{'}(\frac{3}{2}^-)$&0&0   &0   & 1  &  0  &0              &$39\sim31$    &$32\sim111$  &$40\sim25$         &$74\sim258$        &0             &0                 &$185\sim425$         &$1.8\sim10$   &0\\

$\tilde{\Xi}_{c0}(\frac{1}{2}^-)$  & 0 &  0 &  0 & 1  &  1  &$4.6\sim147$   &0             &0            &0                  &0                  &$3.9\sim121$  &0                 &$8.5\sim268$         &$\cdots$ &$\cdots$ \\

$\tilde{\Xi}_{c1}(\frac{1}{2}^-)$  & 0 &  0 &  0 & 1  &  1  &0              &$5.0\sim168$  &$22\sim15$   &$98\sim589$       &$5.4\sim3.1$       &0             &0                 &$130\sim774$         &$0.1\sim0.0$  &0\\

$\tilde{\Xi}_{c1}(\frac{3}{2}^-)$  & 0 &  0 &  0 & 1  &  1  &0              &$19\sim15$    &$31\sim200$  &$20\sim12$        &$140\sim512$       &0             &0                 &$211\sim740$         &$6.9\sim41$    &$0$\\

$\tilde{\Xi}_{c2}(\frac{3}{2}^-)$  & 0 &  0 &  0 & 1  &  1  &$85\sim91$     &$35\sim28$    &$20\sim13$   &$36\sim22$  &$4.9\sim2.8$       &$70\sim75$    &0                 &$252\sim233$         &$0.1\sim0.1$  &0\\

$\tilde{\Xi}_{c2}(\frac{5}{2}^-)$  & 0 &  0 &  0 & 1  &  1  &$85\sim91$     &$16\sim12$    &$31\sim21$   &$16\sim9.9$  &$7.6\sim4.3$       &$70\sim75$    &0                 &$226\sim214$         &$0.5\sim0.4$  &0\\

\hline\hline
\end{tabular}
\label{table12}
\end{table*}
\end{center}
As shown in Tables~\ref{table10} and~\ref{table12}, neither $\Xi_c(3055)$ nor $\Xi_c(3080)$ can be assigned with the $P$-wave excitations. When the facts that the decay mode $\Lambda D$ has been observed while the mode $\Lambda_c K$ has not been unobserved for $\Xi_c(3055)$ and $\Xi_c(3080)$ was taken into account, other assignments except for the $\Xi_{c1}^{'}(\frac{1}{2}^-)$, $\Xi_{c1}(\frac{1}{2}^-)$, $\Xi_{c1}(\frac{3}{2}^-)$ or $\Xi_{c1}^{'}(\frac{3}{2}^-)$ can be excluded. In addition, $\Xi_c(2790)^+$ and $\Xi_c(2815)^+$ have been identified with the candidates of $\Xi_{c1}(\frac{1}{2}^-)$ and $\Xi_{c1}(\frac{3}{2}^-)$. Therefore, only $\Xi_{c1}^{'}(\frac{1}{2}^-)$ and $\Xi_{c1}^{'}(\frac{3}{2}^-)$ are possible assignments for $\Xi_c(3055)$ or $\Xi_c(3080)$. However, in these two assignments, the theoretical predictions of $\Xi_c(3055)$ or $\Xi_c(3080)$ are significantly different from the experimental data both for the decay widths and for the branching ratios.

\subsection{Decays of $\Xi_c(2930)$ and $\Xi_c(3123)$ }

$\Xi_c(2930)$ and $\Xi_c(3123)$ have not well been established in experiment. $\Xi_c(2930)$ was only observed by BaBar in the $\Lambda_c^+ K^-$ invariant mass distribution in an analysis of $B^- \to \Lambda_c^+ \overline{\Lambda}_c^- K^-$~\cite{Aubert:0710.5763} with a width $36\pm7\pm11 $ MeV. $\Xi_c(3123)$ was observed only by BaBar in the $\Sigma_c(2520)^{++} K^- \to \Lambda_c^+ K^- \pi^+$ mass spectrum with a width $4.4\pm3.4\pm1.7 $ MeV with a significance of $3.6$ standard deviations~\cite{Aubert:0710.5763}, while there was no evidence in the $\Lambda_c^+ K^-$ channel.

$\Xi_c(2930)$ and $\Xi_c(3123)$ as possible candidates of $P$-wave excitations, their internal structures and strong decay properties are given in Tables~\ref{table13} and \ref{table14}, respectively.

\begin{center}
  \begin{table*}[htbp]
\caption{Decay widths (MeV) of $ \Xi_c(2930)^+$ as the $P$-wave $\Xi_c$. $\Gamma_{exp}=36\pm13$ MeV, $\mathcal{B}_2=B(\Xi_{c}(2930)^{+} \to \Xi_c^{'0} \pi^{+})/B(\Xi_{c}(2930)^{+} \to \Xi_{c}(2645)^{0}\pi^{+})$. The ranges stand for the results with parameters $\{\gamma, \beta_{\lambda,\rho}\}$ from $\{10.5, 250~\rm {MeV}\}$ to $\{13.4, 400~\rm {MeV}\}$.}
\begin{tabular}{c| ccccc | cccccccccccccccccccccccccccccc}

   \hline\hline
   $ \Xi_{cJ_l} (J^P) $ & $ n_{\lambda} $   & $L_\lambda$         & $ n_{\rho} $    & $L_\rho$      & $S_\rho$
                        &$\Xi_c\pi $        &$\Xi_c^{'}\pi $      &$\Xi_c^{*}\pi$   &$\Lambda_c K $ &$\Gamma_{total}$  & $\mathcal{B}_2 $    \\
   \hline

   $\Xi_{c0}^{'}(\frac{1}{2}^-)$
   & 0 &  1 &  0 & 0  &  1      &$6.3\sim92$    &0              &0             &$12\sim94$      &$19\sim185$   & $\cdots$  \\
   $\Xi_{c1}^{'}(\frac{1}{2}^-)$
   & 0 &  1 &  0 & 0  &  1      &0              &$14\sim63$     &$0.8\sim0.4$  &$0$             &$15\sim64$  &$18\sim161$ \\
   $\Xi_{c1}^{'}(\frac{3}{2}^-)$
   & 0 &  1 &  0 & 0  &  1      &0              &$1.3\sim0.8$   &$18\sim53$    &$0$             &$19\sim54$ &$0.07\sim0.01$\\
   $\Xi_{c2}^{'}(\frac{3}{2}^-)$
   & 0 &  1 &  0 & 0  &  1      &$11\sim8.2$     &$2.4\sim1.4$   &$0.7\sim0.4$  &$6.9\sim4.5$   &$21\sim14$   &$3.4\sim3.9$\\
   $\Xi_{c2}^{'}(\frac{5}{2}^-)$
   & 0 &  1 &  0 & 0  &  1      &$11\sim8.2$    &$1.1\sim0.6$   &$1.1\sim0.6$  &$6.9\sim4.5$    &$20\sim14$  &$1.0\sim1.1$ \\
   $\Xi_{c1}(\frac{1}{2}^-)$
   & 0 &  1 &  0 & 0  &  0      &0              &$7.1\sim32$    &$1.6\sim0.8$  &$0$             &$8.6\sim32$ &$4.6\sim40$ \\
   $\Xi_{c1}(\frac{3}{2}^-)$
   & 0 &  1 &  0 & 0  &  0      &0              &$2.7\sim1.5$   &$9.5\sim27$   &$0$            &$12\sim28$   &$0.3\sim0.06$ \\

$\tilde{\Xi}_{c1}^{'}(\frac{1}{2}^-)$
&0&0   &0   & 1  &  0           &0              &$21\sim95$     &$4.8\sim2.4$  &0              &$26\sim97$    &$4.6\sim40$ \\
$\tilde{\Xi}_{c1}^{'}(\frac{3}{2}^-)$
&0&0   &0   & 1  &  0           &0              &$8.1\sim4.6$    &$28\sim81$   &0              &$37\sim85$  &$0.3\sim0.06$ \\
$\tilde{\Xi}_{c0}(\frac{1}{2}^-)$
& 0 &  0 &  0 & 1  &  1         &$19.2\sim275$  &0              &0             &$36\sim281$    &$56\sim555$   &$\cdots$  \\
$\tilde{\Xi}_{c1}(\frac{1}{2}^-)$
& 0 &  0 &  0 & 1  &  1         &0              &$42\sim190$    &$2.4\sim1.2$  &0              &$45\sim191$   &$18\sim161$ \\
$\tilde{\Xi}_{c1}(\frac{3}{2}^-)$
& 0 &  0 &  0 & 1  &  1         &0              &$4.0\sim2.3$   &$53\sim160$   &0              &$57\sim162$  &$0.07\sim0.01$\\
$\tilde{\Xi}_{c2}(\frac{3}{2}^-)$
& 0 &  0 &  0 & 1  &  1         &$34\sim25$     &$7.3\sim4.2$   &$2.1\sim1.1$  &$21\sim13$   &$64\sim43$    &$3.4\sim3.9$ \\
$\tilde{\Xi}_{c2}(\frac{5}{2}^-)$
& 0 &  0 &  0 & 1  &  1         &$34\sim25$     &$3.2\sim1.6$   &$3.3\sim1.7$  &$21\sim13$   &$61\sim42$    &$1.0\sim1.1$ \\

\hline\hline
\end{tabular}
\label{table13}
\end{table*}
\end{center}

\begin{center}
  \begin{table*}[htbp]
\caption{Decay widths (MeV) of $\Xi_c(3123)^+$ as the $P$-wave $\Xi_c$.
$\Gamma_{exp}=4\pm4$ MeV. The ranges stand for the results with parameters $\{\gamma, \beta_{\lambda,\rho}\}$ from $\{10.5, 250~\rm {MeV}\}$ to $\{13.4, 400~\rm {MeV}\}$.}
\begin{tabular}{c| ccccc | cccccccccccccccccccccccccccccc}

   \hline\hline
   $ \Xi_{cJ_l} (J^P) $ &$ n_{\lambda} $  &$L_\lambda$          &$ n_{\rho} $        &$L_\rho$          &$S_\rho$
                        &$\Xi_c\pi $      &$\Xi_c^{'}\pi $      &$\Xi_c^{*}\pi$      &$\Sigma_c K $     &$\Sigma_c^{*} K $ &$\Lambda_c K $   &$\Lambda D $         &$\Gamma_{total}$   \\

   $\Xi_{c0}^{'}(\frac{1}{2}^-)$   & 0 &  1 &  0 & 0  &  1  &$5.6\sim34$    &0             &0            &0
   &0               &$5.4\sim24$   &$1.9\sim79$       &$13\sim136$          \\

   $\Xi_{c1}^{'}(\frac{1}{2}^-)$   & 0 &  1 &  0 & 0  &  1  &0              &$0.09\sim47$  &$11\sim8.4$  &$13\sim172$                  &$6.3\sim4.1$    &0             &$3.7\sim157$      &$34\sim389$          \\

   $\Xi_{c1}^{'}(\frac{3}{2}^-)$   & 0 &  1 &  0 & 0  &  1  &0              &$8.8\sim7.9$  &$8.5\sim64$  &$13\sim9.0$       &$35\sim185$     &0             &$9.1\sim9.3$      &$74\sim274$          \\

   $\Xi_{c2}^{'}(\frac{3}{2}^-)$   & 0 &  1 &  0 & 0  &  1  &$34\sim42$     &$16\sim14$    &$10\sim7.5$  &$23\sim16$ &$5.7\sim3.7$    &$28\sim35$    &$1.8\sim1.9$      &$117\sim120$         \\

   $\Xi_{c2}^{'}(\frac{5}{2}^-)$   & 0 &  1 &  0 & 0  &  1  &$34\sim42$     &$7.0\sim6.3$  &$16\sim12$   &$10\sim7.2$ &$8.8\sim5.8$    &$28\sim35$    &$29\sim30$        &$132\sim137$         \\

   $\Xi_{c1}(\frac{1}{2}^-)$       & 0 &  1 &  0 & 0  &  0  &0              &$0.05\sim23$  &$22\sim17$   &$6.5\sim86$        &$13\sim8.2$     &0             &$1.9\sim79$       &$43\sim213$          \\

   $\Xi_{c1}(\frac{3}{2}^-)$       & 0 &  1 &  0 & 0  &  0  &0              &$18\sim16$    &$13\sim38$   &$25\sim18$       &$22\sim95$      &0             &$18\sim19$        &$96\sim186$          \\

$\tilde{\Xi}_{c1}^{'}(\frac{1}{2}^-)$&0&0   &0   & 1  &  0  &0              &$0.1\sim70$   &$66\sim50$   &$20\sim259$        &$38\sim25$         &0             &0                 &$124\sim404$         \\

$\tilde{\Xi}_{c1}^{'}(\frac{3}{2}^-)$&0&0   &0   & 1  &  0  &0              &$53\sim47$    &$38\sim114$  &$76\sim54$         &$66\sim286$        &0             &0                 &$232\sim502$         \\

$\tilde{\Xi}_{c0}(\frac{1}{2}^-)$  & 0 &  0 &  0 & 1  &  1  &$17\sim101$    &0             &0            &0                  &0                  &$16\sim72$    &0                 &$33\sim173$          \\

$\tilde{\Xi}_{c1}(\frac{1}{2}^-)$  & 0 &  0 &  0 & 1  &  1  &0              &$0.3\sim141$  &$33\sim25$   &$39\sim517$       &$19\sim12$         &0             &0                 &$92\sim696$          \\

$\tilde{\Xi}_{c1}(\frac{3}{2}^-)$  & 0 &  0 &  0 & 1  &  1  &0              &$26\sim24$    &$25\sim191$  &$38\sim27$        &$104\sim554$       &0             &0                 &$194\sim795$         \\

$\tilde{\Xi}_{c2}(\frac{3}{2}^-)$  & 0 &  0 &  0 & 1  &  1  &$101\sim124$   &$47\sim43$    &$30\sim23$   &$68\sim49$  &$17\sim11$         &$83\sim106$   &0                 &$347\sim355$         \\

$\tilde{\Xi}_{c2}(\frac{5}{2}^-)$  & 0 &  0 &  0 & 1  &  1  &$101\sim124$   &$21\sim19$    &$47\sim35$   &$30\sim22$  &$26\sim17$         &$83\sim106$   &0                 &$309\sim323$         \\

\hline\hline
\end{tabular}
\label{table14}
\end{table*}
\end{center}

As seen in Table~\ref{table13}, the assignment of $\Xi_c(2930)$ with the $\Xi_{c1}^{'}(\frac{1}{2}^-)$, $\Xi_{c1}^{'}(\frac{3}{2}^-)$, $\Xi_{c1}(\frac{1}{2}^-)$, $\Xi_{c1}(\frac{3}{2}^-)$, $\tilde{\Xi}_{c1}^{'}(\frac{1}{2}^-)$, $\tilde{\Xi}_{c1}^{'}(\frac{3}{2}^-)$, $\tilde{\Xi}_{c1}(\frac{1}{2}^-)$, or $\tilde{\Xi}_{c1}(\frac{3}{2}^-)$ is impossible for the forbidden $\Lambda_c^+ K^-$ decay mode. Other possibilities can not be excluded for lack of experimental information.

Besides, we notice that if $\Xi_c(2930)$ is assigned with the $\Xi_{c0}^{'}(\frac{1}{2}^-)$ or $\tilde{\Xi}_{c0}(\frac{1}{2}^-)$, the $ \Xi_c^{'} \pi$ and $ \Xi_c^{*} \pi$ decay modes will be forbidden. If $\Xi_c(2930)$ is assigned with $\Xi_{c2}^{'}(\frac{3}{2}^-)$, $\Xi_{c2}^{'}(\frac{5}{2}^-)$, $\tilde{\Xi}_{c2}(\frac{3}{2}^-)$ or $\tilde{\Xi}_{c2}(\frac{5}{2}^-)$,
the ratio $\mathcal{B}_2=B(\Xi_{c}(2930)^{+} \to \Xi_c^{'0} \pi^{+})/B(\Xi_{c}(2930)^{+} \to \Xi_{c}(2645)^{0}\pi^{+})$ will be an important signal for the understanding of its internal structure. In order to classify the $\Xi_c(2930)$ state, more experimental information is required.

From Table~\ref{table14}, $\Xi_c(3123)$ seems impossible to be identified with the $P$-wave excitations for the large widths, which significantly disagree with a narrow measured width $\Gamma_{exp}=4\pm4$ MeV.

\section{Conclusions and discussions\label{Sec: summary}}

In this work, the strong decay properties of $\Xi_c(2790)$, $\Xi_c(2815)$, $\Xi_c(2980)$, $\Xi_c(3055)$, $\Xi_c(3080)$, $\Xi_c(2930)$, and $\Xi_c(3123)$ as possible $P$-wave excited $\Xi_c$ baryons have been systematically studied in a $^3P_0$ model. Possible configurations and assignments of $\Xi_c(2790)$, $\Xi_c(2815)$, $\Xi_c(2930)$, $\Xi_c(2980)$, $\Xi_c(3055)$, $\Xi_c(3080)$, and $\Xi_c(3123)$ have been analyzed in comparison with experimental data.

Our main conclusions are as follows:

$1$, $\Xi_c(2790)$ and $\Xi_c(2815)$ are very possibly the $P$-wave $\Xi_{c1}(1/2^-)$ and $\Xi_{c1}(3/2^-)$ $\Xi_c$, respectively. The predicted strong decay width ($\Gamma_{total}=7.2\sim16.6$ MeV) of $\Xi_c(2790)$ is well consistent with the measured one ($\Gamma_{\Xi_c(2790)^+ }=8.9\pm0.6\pm0.8 $ MeV). For $\Xi_c(2815)$, both the strong decay width ($\Gamma_{total}=4.8\sim9.4 $ MeV) and the ratio $\mathcal{B}_1=B(\Xi_{c}(2815)^{+} \to \Xi_c^{'0} \pi^{+})/B(\Xi_{c}(2815)^{+} \to \Xi_{c}(2645)^{0}\pi^{+})=7.2\sim1.7 \%$ are compatible with the experimentally measured $\Gamma_{\Xi_c(2815)^+ }=2.43\pm0.20\pm0.17 $ MeV and $\mathcal{B}_1= 11 \%$.

$2$, $\Xi_c(2980)$ may be the $P$-wave excited $\Xi_{c1}^{'}(\frac{1}{2}^-)$. In this assignment, the ratios,  $\mathcal{B}_2^{'}=B(\Xi_{c}(2980)^{+} \to \Xi_c^{'0} \pi^{+})/B(\Xi_{c}(2815)^{+} \to \Xi_{c}(2645)^{0}\pi^{+},\Xi_{c}(2645)^{0} \to \Xi_c^{+}\pi^{-})=45\sim81\%$, $\mathcal{B}_2=B(\Xi_{c}(2980)^{+} \to \Xi_c^{'0} \pi^{+})/B(\Xi_{c}(2980)^{+} \to \Xi_{c}(2645)^{0}\pi^{+})=7.0\sim77$ and $\mathcal{B}_3=B(\Xi_{c}(2980)^{+} \to \Sigma_c^{++} K^-)/B(\Xi_{c}(2980)^{+} \to \Xi_{c}(2645)^{0}\pi^{+})=36\sim144$ were obtained. The measurements of these ratios in the future will be helpful for the understanding of this $\Xi_c$.

$3$, $\Xi_c(2930)$ has not been well established in experiment. Our numerical results suggest that $\Xi_c(2930)$ could be assigned as the $P$-wave $\Xi_{c0}^{'}(\frac{1}{2}^-)$ or $\tilde{\Xi}_{c0}(\frac{1}{2}^-)$. In these two assignments, the $ \Xi_c^{'} \pi$ and $ \Xi_c^{*} \pi$ decay modes are forbidden. $\Xi_c(2930)$ is also possible a $\Xi_{c2}^{'}(\frac{3}{2}^-)$, $\Xi_{c2}^{'}(\frac{5}{2}^-)$, $\tilde{\Xi}_{c2}(\frac{3}{2}^-)$ or $\tilde{\Xi}_{c2}(\frac{5}{2}^-)$. In these assignments, the ratios $\mathcal{B}_2=B(\Xi_{c}(2930)^{+} \to \Xi_c^{'0} \pi^{+})/B(\Xi_{c}(2930)^{+} \to \Xi_{c}(2645)^{0}\pi^{+})$ are important for the understanding of its internal structure. More experimental information is required to identify the $\Xi_c(2930)$.

$4$, $\Xi_c(3055)$, $\Xi_c(3080)$ and $\Xi_c(3123)$ could be excluded as $P$-wave excitations of $\Xi_c$.

As a phenomenological model to study OZI-allowed hadronic decays of hadron, there are some uncertainties from the parameters in the $^3P_0$ model. How to fix the parameters such as $\gamma$ and $\beta_{\lambda,\rho}$ is an important topic in the future. Other possible assignments of these excited $\Xi_c$ have not explored in this paper. In addition, it is necessary to study systemically the hadronic decay of these excited $\Xi_c$ in much more other models as a crosscheck.

\begin{acknowledgments}
 This work is supported by National Natural Science Foundation of China under the grants No: 11475111 and No: 11075102.
\end{acknowledgments}

\end{document}